\documentclass[11pt]{article}
\pdfoutput=1
\usepackage{latexsym}
\usepackage{amssymb}
\usepackage{epsf}
\usepackage{amsmath}
\usepackage{comment}
\usepackage{bm}
\usepackage{graphicx}% Include figure files

\begin{document}
\pagestyle{empty}
\renewcommand{\thefootnote}{\fnsymbol{footnote}}

\vspace*{1cm}

\begin{center}

{\bf \LARGE Magnetization Dynamics in  
Holographic \\ Ferromagnets: \\
\vspace*{0.3cm}
Landau-Lifshitz Equation from Yang-Mills Fields} 
\\

\vspace*{2cm}
{\large 
Naoto Yokoi$^{1, 2}$\footnote{yokoi@spin.t.u-tokyo.ac.jp}, 
Koji Sato$^{2}$\footnote{koji.sato.c5@tohoku.ac.jp}, 
and Eiji Saitoh$^{1, 2, 3, 4, 5}$\footnote{eizi@ap.t.u-tokyo.ac.jp}} \\
\vspace*{1.5cm}

{\it $^{1}$Department of Applied Physics, The University of Tokyo, Tokyo 113-8656, Japan \\
$^{2}$Institute for Materials Research, Tohoku University, Sendai 980-8577, Japan \\
$^{3}$Advanced Institute for Materials Research, Tohoku University, 
Sendai 980-8577, Japan \\
$^{4}$Center for Spintronics Research Network, Tohoku University, Sendai 980-8577, Japan \\
$^{5}$Advanced Science Research Center, Japan Atomic Energy Agency, Tokai 319-1195, Japan}
\end{center}

\vspace*{3.5cm}

\begin{abstract}
{\normalsize We introduce a new approach to understand magnetization dynamics 
in ferromagnets based on the holographic realization of ferromagnets. 
A Landau-Lifshitz equation describing the magnetization dynamics is  
derived from a Yang-Mills equation in the dual gravitational theory,  
and temperature dependences of the spin-wave stiffness and 
spin transfer torque appearing in the holographic Landau-Lifshitz equation  
are investigated by the holographic approach.    
The results are consistent with the known properties of magnetization dynamics 
in ferromagnets with conduction electrons.  
}
\end{abstract} 

%%%%%%%%%%%%%%%%%%%%%%%%%%%%%%%%%%%%%%%%%%%%%%%%%%%%%%%%%%%%%%%%%%%%%%%%%%%%
\newpage
\baselineskip=18pt
\setcounter{page}{2}
\pagestyle{plain}
\baselineskip=18pt

\renewcommand{\thefootnote}{\arabic{footnote}}
\setcounter{footnote}{0}

\section{Introduction}
The Landau-Lifshitz equation \cite{LLequation} is the fundamental equation 
for describing the dynamics of magnetization (density of magnetic moments) 
in various magnetic materials. 
It has been also playing a fundamental role 
in the development of modern spintronics \cite{SpinCurrent}:  
For instance, 
its extension to the coupled systems of localized magnetic moments 
and conduction electrons has   
led to the concepts of spin transfer torque \cite{Berger, Slonczewski} and spin pumping \cite{Tserkovnyak}. 
So far, the symmetries and reciprocity in electronic systems have been 
the guiding principles to develop such extensions. 
In this article, we introduce another guiding principle to explore 
the new extensions and magnetization dynamics on the basis of the holographic duality.

The holographic duality is the duality between the quantum many body system defined in 
$d$-dimensional space-time and the gravitational theory (with some matter fields) which lives 
in $(d+1)$-dimensional space-time 
\cite{Maldacena:1997re, Gubser:1998bc, Witten:1998qj}.\footnote{See \cite{AdSCMT} for 
a recent review on the applications of the holographic duality to condensed matter physics.}   
We constructed a holographic dual model of three-dimensional ferromagnetic systems, which 
exhibits the ferromagnetic phase transition with spontaneous magnetization and 
the consistent magnetic properties 
at low temperatures \cite{Yokoi:2015qba}.\footnote{Other holographic approaches to 
ferromagnetic systems have been also discussed in \cite{Iqbal:2010eh, Cai:2014oca, Cai:2015jta}.} 
In the holographic duality, finite temperature effect in ferromagnetic systems can be 
incorporated as the geometrical effect of black holes 
in higher dimensional bulk gravity, and 
the Wick rotation at finite temperatures is not required for 
the analysis in the dual gravitational theory. Thus, the novel analysis for real-time dynamics  
of quantum many body systems in nonequilibrium situations can be performed 
using the holographic approach (for a review, see \cite{Hubeny:2010ry}). 
In addition, the holographic duality is known to be a strong-weak duality, 
which relates strongly correlated quantum systems to classical gravitational theories.  
From these viewpoints, the holographic approach can provide 
new useful tools to analyze nonequilibrium and nonlinear dynamics  
of magnetization in ferromagnets.           

In ferromagnets, spin currents are generated by magnetization dynamics. 
From the holographic dictionary between the quantities of ferromagnets and gravitational theory 
\cite{Yokoi:2015qba},   
the spin currents in ferromagnets 
correspond to the $SU(2)$ gauge fields in the dual gravitational theory. 
This correspondence indicates that 
the dynamics of spin currents, consequently the dynamics of magnetization, can be 
described by the Yang-Mills equation for $SU(2)$ gauge fields \cite{Yang:1954ek} 
in the holographic dual theory. 
In the following, we derive a Landau-Lifshitz equation for magnetization dynamics 
from the Yang-Mills equation within the holographic realization of ferromagnets. 
This derivation can provide novel perspectives for magnetization dynamics from the non-abelian gauge theory. 

This article is organized as follows. 
In Section 2, we summarize the results of the magnetic properties  
obtained from the holographic realization of ferromagnets in thermodynamic equilibrium. 
An extension to nonequilibrium situation including the fluctuations of magnetization and 
spin currents is discussed in the dual gravitational theory, and the holographic equation of 
magnetization dynamics is derived in Section 3. In Section 4, temperature dependences of 
the parameters in the resulting holographic equation are investigated by numerical calculations. 
Finally, we summarize the results in Section 5.

\section{Holographic Dual Model of Ferromagnets}
We begin with a brief summary on the holographic dual model 
of ferromagnets \cite{Yokoi:2015qba}.
The dual model is the five-dimensional gravitational theory with an $SU(2)$ gauge field $A_{M}^{a}$ and  
a $U(1)$ gauge field $B_{M}$, whose action is given by 
\begin{eqnarray}
S = \int\!\sqrt{- g}\, d^5 x&&\hspace*{-0.6cm}\left[\,\frac{1}{2 \kappa^2}\left( R - 2 \Lambda \right) 
- \frac{1}{4 e^{2}} G_{M N} G^{M N} 
- \frac{1}{4 g^{2}} F_{M N}^{a} F^{a\, M N} \right. \nonumber \\
&&\hspace*{-0.3cm}\left. - \frac{1}{2} \left(D_{M} \phi^{a}\right)^2 - V(|\phi|)\,\right]\,.
\label{eq: holographic action}
\end{eqnarray}
Here, $R$ is the scalar curvature of space-time, and the field strength is defined by $F_{M N}^{a} = \partial_{M} A_{N}^{a} - \partial_{N} A_{M}^{a} + \epsilon^{a b c}A_{M}^{b}A_{N}^{c}$ and $G_{M N} = \partial_{M} B_{N} - \partial_{N} B_{M}$, respectively. The index $a$ labels spin directions in the $SU(2)$ 
space ($a = 1 \sim 3$), the index $M$ labels 
space-time directions in five dimensions ($M, N = 0 \sim 4$), 
and $\epsilon^{a b c}$ is a totally anti-symmetric tensor with $\epsilon^{1 2 3} = 1$. 
The model also includes a triplet scalar field $\phi^{a}$ with the covariant derivative 
$D_{M} \phi^{a} = \partial_{M} \phi^{a} + \epsilon^{a b c} A_{M}^{b} \phi^{c}$, and 
the $SU(2)$-invariant scalar potential $V(|\phi|)$ with 
the norm $|\phi|^2 = \sum_{a=1}^{3}(\phi^{a})^{2}$. Note that the scalar field is neutral 
under the $U(1)$ gauge transformation.   
In order to guarantee asymptotic Anti-de Sitter (AdS) backgrounds, the negative 
cosmological constant $\Lambda = - 6/\ell^2$ is introduced. 
The field-operator correspondence in the holographic duality \cite{Gubser:1998bc, Witten:1998qj} 
leads to the following holographic dictionary between the fields of the dual gravitational theory 
and the physical quantities of ferromagnets: 
\begin{table}[h]
\begin{center}
\begin{tabular}{ccc}
Dual gravity & & Ferromagnet \\ \hline
Scalar field~ $\phi^{a}$ & $\Longleftrightarrow$ & Magnetization~ $M^{a}$ \\
$SU(2)$ gauge field~ $A_{M}^{a}$ & $\Longleftrightarrow$ & Spin current~ $J_{s\, \mu}^{a}$ \\
$U(1)$ gauge field~ $B_M$ & $\Longleftrightarrow$ & Charge current~ $J_\mu$ \\
Metric~ $g_{M N}$ & $\Longleftrightarrow$ & Stress tensor~ $T_{\mu\nu}$ \\ \hline
\end{tabular}
\caption{Holographic dictionary between the dual gravitational theory and 
ferromagnets.}
\end{center}
\end{table}

\subsection{Black Hole as Heat Bath}
In order to establish the holographic dictionary, thermodynamical properties 
of the physical quantities of ferromagnets should be calculated in the dual gravitational theory. 
In Ref. \cite{Yokoi:2015qba}, the temperature dependences of magnetic quantities and 
the behavior of ferromagnetic phase transition are thoroughly discussed. 
In the context of the holographic duality, finite temperature effects in 
the ferromagnets can be incorporated by introducing the black holes into 
the dual gravitational theory as the background space-time.    
Indeed, the dual gravitational theory has 
the charged black hole solution which is a solution to the Einstein, Yang-Mills, and 
Maxwell equations derived from the action (\ref{eq: holographic action}):
\begin{eqnarray}
&&R_{M N} + \left( \Lambda - \frac{1}{2} R \right) g_{M N} = \frac{\kappa^2}{2 e^2} \left( 2 G_{K M} G^{K}{}_{N} - \frac{1}{2} G_{K L} G^{K L} g_{M N} \right) \nonumber \\
&&\hspace*{4.2cm}+\,\frac{\kappa^2}{2 g^2} \left( 2 F^{a}_{K M} F^{a K}{}_{N} - \frac{1}{2} F_{a K L} 
F^{a K L} g_{M N} \right) , \label{eq: Einstein equation}\\
&&\nabla_{M} F^{a M N} + \epsilon^{a b c} A^{b}_{M} F^{c M N} = 0 , \qquad \nabla_{M} G^{M N} = 0 , 
\end{eqnarray}   
where $\nabla_{M}$ is the covariant derivative for the affine connection, and 
the space-time indices $M, N$ are raised or lowered by the bulk metric $g_{M N}$.  
Here, we neglect the contribution from the scalar field and set $\phi^{a}=0$ for the background. 
The metric of the black hole\footnote{This type of non-abelian black holes has been 
discussed in the context of the holographic duality, in the literature such as 
\cite{Iqbal:2010eh, Herzog:2009ci}.} 
is given by
\begin{equation}
ds^2 = g_{M N}\,dx^{M} dx^{N} = \frac{r^2}{\ell^2} \left(-f(r)\,dt^2 + dx^2 + dy^2 + dz^2 \right) 
+ \frac{\ell^2}{f(r)} \frac{dr^2}{r^2}\,, \label{eq: BH metric in r}
\end{equation}
with the radial function, 
\begin{equation}
f(r)=1 - (1 + Q^2) \left(\frac{r_{H}}{r}\right)^{4} 
+ Q^{2} \left(\frac{r_{H}}{r}\right)^{6}\,. \label{eq: horizon in r}
\end{equation}
Here, we define the parameter $Q$:
\begin{equation}
Q^2 = \frac{2 \kappa^2}{3} \left(\frac{\mu_{e}^2}{e^2} + \frac{\mu_{s}^2}{g^2} \right) .
\label{eq: chemical potential}
\end{equation} 
The $U(1)$ charge $\mu_{e}$ and $SU(2)$ charge $\mu_{s}$ of the black hole 
are supported by the time components of the gauge fields, 
\begin{eqnarray}
B_{0} = \mu_{e}\left(\frac{r_{H}}{\ell}\right)\left(1 - \frac{r_{H}^2}{r^2}\right)\qquad \textrm{and}\qquad 
A_{0}^{3} = \mu_{s}\left(\frac{r_{H}}{\ell}\right)\left(1 - \frac{r_{H}^2}{r^2}\right) .  
\label{eq: time component of gauge field}
\end{eqnarray}
Note that the black hole solution (\ref{eq: BH metric in r}) is asymptotically AdS at $r\rightarrow\infty$, 
and has the (outer) horizon $r=r_{H}$. 

For the following discussion, we make a coordinate change of 
the radial coordinate $r$ into $u$ by $u = 1/r$, and the black hole metric becomes 
\begin{eqnarray}
ds^2=\frac{1}{u^2} \left(- f(u)\,dt^2 + dx^2 + dy^2 + dz^2 + \frac{du^2}{f(u)} \right) ,
\label{eq: BH metric in u}
\end{eqnarray}
and the transformed function $f(u)$ is given by 
\begin{eqnarray}
f(u) = 1 - \left(1+Q^2\right) u^4 + Q^{2} u^{6} ,
\label{eq: horizon in u}
\end{eqnarray}
where we have set the coupling parameters $e = g = 1$  
and the black hole parameters $r_{H} = \ell = 1$, for simplicity. 

In the holographic dual model, the black hole (\ref{eq: BH metric in u}) 
plays the role of the heat bath; due to the Hawking radiation, the black hole temperature is given by     
\begin{eqnarray}
T = \frac{2 - Q^2}{2 \pi} ,
\label{eq: Hawking temperature}
\end{eqnarray}
and the calculations on the black hole background lead to the thermodynamical properties 
of the corresponding ferromagnet. 
Since we focus only on the dynamics of magnetization and spin current, 
the background space-time is fixed to be the black hole metric (\ref{eq: BH metric in u}) in the following.  

\subsection{Thermodynamics of Ferromagnets from Scalar Dynamics on Charged Black Hole} 
In order to investigate the thermodynamics of magnetization, we examine the equation of motion 
for the scalar field $\phi^{a}$, which is also derived from the action (\ref{eq: holographic action}):
\begin{eqnarray}
\frac{1}{\sqrt{-g}} \partial_{M}\left( \sqrt{- g}\, D^{M} \phi^{a} \right) + 
\varepsilon^{a b c} A_{M}^{b} D^{M}\phi^{c} = \frac{\partial V}{\partial \phi^{a}} .
\label{eq: scalar equation}
\end{eqnarray} 
Here, we consider a static and homogeneous solution in the boundary coordinates,  
$x^{\mu} = (t, x^{1},x^{2},x^{3})$, which corresponds to 
the homogeneous magnetization in ferromagnets.  
Without loss of generality, the ansatz for such a scalar field, which is  
invariant under the translations on the boundary, is given by 
\begin{eqnarray}
\phi^{1} = \phi^{2} = 0, \quad \phi^{3} = \Phi(u) \neq 0 .
\label{eq: static scalar ansatz}
\end{eqnarray}
\begin{comment}
This ansatz implies that the cross terms identically vanish: 
\begin{eqnarray}
\varepsilon^{a b c} A_{M}^{b} A_{N}^{c} = \varepsilon^{a b c} A_{M}^{b} \partial_{N} \phi^{c} = 
\varepsilon^{a b c} A_{M}^{b} \phi^{c} = \varepsilon^{a b c} \phi^{b} \partial_{N}\phi^{c}  = 0 .
\end{eqnarray}
In particular, from this property, the currents (\ref{eq: YM current}) identically vanish. 
\end{comment}
Inserting this ansatz, the metric (\ref{eq: BH metric in u}), and 
the gauge fields (\ref{eq: time component of gauge field}) into the equation (\ref{eq: scalar equation}),   
we obtain the following equation for $\Phi(u)$: 
\begin{eqnarray}
u^{2} f(u)\, \frac{d^{2} \Phi}{d u^{2}} + \left(u^{2}\, \frac{d f(u)}{d u} - 3\,u\,f(u) \right) \frac{d \Phi}{d u} = 
\frac{\partial V}{\partial \Phi} . 
\label{eq: static equation for phi}
\end{eqnarray} 
This equation governs the thermodynamics of magnetization in the dual gravitational theory. 
We can analyze the solution to this equation numerically with a simple quartic potential 
$V(|\phi|) = \lambda \left(|\phi|^2-m^2/\lambda\right)^2\!/4$, and 
the asymptotic behavior of the numerical solution near the boundary $u \sim 0$ (or $r \sim \infty$) 
is obtained: 
\begin{eqnarray}
\Phi(u)\, \simeq\, H_{0}\, u^{\Delta_{-}} + M(T)\,u^{\Delta_{+}} \qquad 
\left(\Delta_{\pm} = 2 \pm \sqrt{4-m^{2}}\right) .
\label{eq: static solution for phi}
\end{eqnarray}
According to the standard recipe in the holographic duality \cite{Balasubramanian:1998de, Klebanov:1999tb},
the coefficients $H_{0}$ and $M(T)$ in the asymptotic expansion correspond to 
an external magnetic field and a magnetization at temperature $T$ (under $H_{0}$), respectively. 
In Ref.\,\cite{Yokoi:2015qba}, the resulting temperature dependences of magnetization, 
magnetic susceptibility, and specific heat have been shown to 
reproduce the ferromagnetic phase transition in the mean field theory. 
Furthermore, the temperature dependences at low temperatures 
are also consistent with the existence of the spin wave excitations (magnons) and 
conduction electrons in low-temperature ferromagnets.   

For later convenience, we also comment on the solutions for the gauge fields. 
Assuming the translational and rotational invariance on the boundary, 
equilibrium solutions for the gauge fields are given by the following form:
\begin{eqnarray}
B_{0} = b(u) \quad \textrm{and} \quad A_{0}^{3} = a^{3}(u) ,
\label{eq: static gauge ansatz}
\end{eqnarray}
where all the other components vanish. 
Inserting this ansatz and  (\ref{eq: static scalar ansatz}), 
the Maxwell and Yang-Mills equations on the black hole are 
reduced to the following simple forms:
\begin{eqnarray}
\frac{d}{d u} \left(\frac{1}{u}\,\frac{d b}{d u}\right) = 0 \quad \textrm{and} \quad 
\frac{d}{d u} \left(\frac{1}{u}\,\frac{d a^{3}}{d u}\right) = 0 .
\label{eq: Maxwell equation}
\end{eqnarray}
The general solutions are given by the forms (\ref{eq: time component of gauge field}) 
in terms of $u$,  
\begin{eqnarray}
b(u) = \mu_{e} \left(1 - u^2\right)\quad \textrm{and}\quad a^{3}(u) = \mu_{s} \left(1 - u^2\right) .
\label{eq: static solution for time component}
\end{eqnarray}
Here, we impose the boundary conditions $B_{0} = 0$ and $A_{0}^{3} = 0$ at the horizon ($u=1$), 
which guarantee the regularity of the gauge fields on the horizon.  
The remaining integral constants, $\mu_{e}$ and $\mu_{s}$, correspond respectively to 
the electrochemical potential of underlying electrons  
and the spin chemical potential (or spin voltage), through the holographic dictionary.

To summarize, the solutions (\ref{eq: static solution for phi}) and (\ref{eq: static solution for time component}) 
on the charged black hole describe the thermodynamical property of 
the holographic dual ferromagnets in the equilibrium.

\section{Magnetization Dynamics in Holographic Ferromagnets}
In this section, we extend the holographic analysis in the equilibrium, summarized in the previous section, 
to more general situations including the dynamics of magnetization and spin currents. 
In order to discuss the dynamics of magnetization and spin currents, 
the static and homogeneous ansatze for the scalar field (\ref{eq: static scalar ansatz}) and 
the gauge fields (\ref{eq: static gauge ansatz}) need to be generalized. 
Here, we focus on the dynamics with the long wave length in 
the ordered phase (symmetry broken phase) below the Curie temperature, 
where various phenomena in modern spintronics are intensively studied.  

\subsection{Generalized Ansatz and Effective Equations of Motion} 
For the scalar field, following the standard derivation of the equation for 
magnetization dynamics, 
we consider the generalized ansatz for the scalar field as a factorized form:
\begin{eqnarray}
\phi^{a}(u, t, x) = \Phi(u)\, n^{a}(t,x)\, \quad \textrm{with} \quad \sum_{a=1}^{3} n^{a}n^{a} = 1,
\label{eq: magnetization ansatz}
\end{eqnarray}   
where $\Phi(u)$ is a solution of the equation (\ref{eq: static equation for phi}) 
with the asymtotic behavior (\ref{eq: static solution for phi}). 
Note that, since we focus only on the dynamics of spontaneous magnetization, 
we fix $H_{0} = 0$ throughout this article.  
In this ansatz, $n^{a}(t, x)$ corresponds to the (local) direction of magnetization 
in ferromagnets.   

In ferromagnetic systems, the magnetization dynamics generates 
various dynamics of spin currents \cite{SpinCurrent}. 
In the holographic dual theory, the scalar dynamics is also expected to induce 
the dynamics of the corresponding $SU(2)$ gauge field,    
and thus we generalize the static and homogeneous ansatz for the $SU(2)$ gauge fields to 
the following factorized forms:
\begin{eqnarray}
A_{0}^{\parallel}(u, t, x) &=& (1 - u^2)~ a_{0}^{\parallel}(t, x), \nonumber \\
A_{0}^{\perp}(u, t, x) &=& (1 - u^2)~ a_{0}^{\perp}(t, x), \nonumber \\
A_{i}^{\parallel}(u, t, x) &=& G^{\parallel}(u)~ a_{i}^{\parallel}(t, x), \nonumber \\
A_{i}^{\perp}(u, t, x) &=& G^{\perp}(u)~ a_{i}^{\perp}(t, x) \qquad  \left(i = 1 \sim 3\right) ,
\label{eq: gauge ansatz}
\end{eqnarray}
where we set the radial component $A^{a}_{u} \equiv 0$ by using the gauge degrees of freedom. 
Due to the nontrivial scalar solution $\Phi(u)$, corresponding to the spontaneous magnetization, 
the $SU(2)$ gauge symmetry is broken to $U(1)$. The gauge fields can be correspondingly 
decomposed into an unbroken component $A_{\mu}^{\parallel}$ 
and two broken components $A_{\mu}^{\perp}$, 
which are defined by $A_{\mu}^{\parallel} \propto n^{a}$ and $n \cdot A_{\mu}^{\perp}=0$, respectively.  
As in the case of the static solutions, the time components of gauge fields should satisfy 
the horizon boundary condition, $A_{0}^{a} = 0$ at $u=1$, for the regularity. 
Although the spatial components $A_{i}^{a}$ are not required to vanish on the horizon, 
the regularity (or finiteness) at the horizon is required.   
The asymptotic solutions to the linearized Yang-Mills equation near the boundary ($u\sim0$)  
give the asymptotic expansions for the radial functions $G^{\parallel}(u)$, and $G^{\perp}(u)$,
\begin{eqnarray} 
G^{\parallel}(u) &=& 1 -  \sigma_{s}^{\parallel}\, u^2 + {\cal O}(u^4), \nonumber \\
G^{\perp}(u) &=& 1 + \sigma_{s}^{\perp}\, u^2 + {\cal O}(u^4).
\label{eq: radial solution for A}
\end{eqnarray}
We discuss the concrete numerical solutions of $G^{a}(u)$ and their physical implications 
in the next section. 

Since the scalar field $\phi^{a}$ does not have the $U(1)$ charge, 
the fluctuation (or dynamics) of $\phi^{a}$ does not induce further dynamics for the 
$U(1)$ gauge field, which implies the solution for $B_{\mu}$ in 
(\ref{eq: time component of gauge field}) is unchanged, and we can neglect the dynamics of $B_{\mu}$.  

At first, we consider the equation of motion for the scalar field $\phi^{a}$. 
Inserting the generalized ansatz (\ref{eq: magnetization ansatz}) into the equation (\ref{eq: scalar equation}), 
we obtain the following equation for $n^{a}$:
\begin{eqnarray}
\left[ u^{5} \partial_{u}\left(u^{-3} f(u)\,\partial_{u} \Phi \right) 
- \frac{\partial V}{\partial \Phi} \right] n^{a} 
= \left[ \frac{u^{2}}{f(u)} D_{t} D_{t} n^{a} - u^{2} D_{i} D_{i} n^{a} \right] \Phi .
\label{eq: factorized equation}
\end{eqnarray} 
Here, we have used the gauge condition $A_{u}^{a} = 0$, and 
the gauge covariant derivative is defined as $D_{\mu} n^{a} = \partial_{\mu} n^{a} + 
\varepsilon^{a b c} A_{\mu}^{b} n^{c}$. 
The left-hand side of the equation (\ref{eq: factorized equation}) is proportional to 
the equation (\ref{eq: static equation for phi}), and thus vanishes for the solution $\Phi(u)$. 
Since $\Phi(u)$ is a non-trivial solution, which is not identically zero, we have the effective equation of 
motion for $n^{a}$:
\begin{eqnarray}
f^{-1} D_{t} D_{t}\,n^{a} - D_{i} D_{i}\,n^{a} = 0 .
\label{eq: equation for na}
\end{eqnarray}    

Next, the equation of motion for the gauge fields is considered.  
The Yang-Mills equation for the $SU(2)$ gauge field $A_{M}^{a}$ is derived by 
the variation of the holographic action (\ref{eq: holographic action}) and given by   
\begin{eqnarray}
\frac{1}{\sqrt{-g}} \partial_{N}\left( \sqrt{-g}\, F^{N M\,a} \right) + 
\epsilon^{a b c} A^{b}_{N} F^{N M\,c} = J^{M\,a} ,
\label{eq: YM equation}
\end{eqnarray}
where the $SU(2)$ current is defined as
\begin{eqnarray}
J_{M}^{a} = \varepsilon^{a b c} \phi^{b} D_{M} \phi^{c} = \varepsilon^{a b c} \phi^{b} 
\left(\partial_{M} \phi^{c} + \varepsilon^{c d e} A_{M}^{d} \phi^{e} \right) .
\label{eq: YM current}
\end{eqnarray}
Unlike the static case, the generalized ansatz (\ref{eq: magnetization ansatz}) 
and (\ref{eq: gauge ansatz}) give the non-vanishing currents:
\begin{eqnarray}
J_{\mu}^{a} 
= \Phi^{2} \left(\varepsilon^{a b c}\, n^{b} \partial_{\mu} n^{c} + 
\varepsilon^{a b c} \varepsilon^{c d e}\, n^{b} a_{\mu}^{d} n^{e} + {\cal O}\left(u^2\right) \right) .
\label{eq: generalized current}
\end{eqnarray}
Note that the radial component of the currents still vanishes, $J_{u}^{a} \equiv 0$, 
due to the gauge fixing condition $A_{u}^{a} \equiv 0$. 
With this current, we can explicitly write down the Yang-Mills equations 
on the charged black hole (\ref{eq: BH metric in u}), in the boundary direction\,: 
\begin{eqnarray}
J_{0}^{a} &=& u^{3} f\, \partial_{u} \left(u^{-1} F_{u 0}^{a}\right) +  
u^{2} \left( D_{i} F_{i 0}^{a} \right), \label{eq: generalized time YM equation}\\
J_{i}^{a} &=& u^{3}\, \partial_{u} \left(u^{-1} f\, F_{u i}^{a}\right) 
- u^{2} f^{-1} \left( D_{0} F_{0 i}^{a} \right) + u^{2} \left( D_{j} F_{j i}^{a} \right) , 
\label{eq: generalized space YM equation}
\end{eqnarray}
where the gauge covariant derivative for the field strength is defined as $D_{\mu} F_{\nu \rho}^{a} = \partial_{\mu} F_{\nu \rho}^{a} + \varepsilon^{a b c} A_{\mu}^{b} F_{\nu \rho}^{c}$.
Inserting the ansatz (\ref{eq: gauge ansatz}), the Yang-Mills equations give the equations for 
$n^{a}$ and $a_{\mu}^{a}$.  
In summary, using the generalized ansatze, we have obtained the coupled equations of 
motion for $n^{a}$ and $a_{\mu}^{a}$, (\ref{eq: equation for na}), (\ref{eq: generalized time YM equation}),  
and (\ref{eq: generalized space YM equation}).

\subsection{Landau-Lifshitz Equation from Yang-Mills Equation}
Since it is difficult to find the general solutions for the coupled non-linear partial differential equations,  
we seek simple trial solutions for $n^{a}$ and $a_{\mu}^{a}$ to obtain the effective equations of motion. 
At first, instead of looking for general solutions to the equation (\ref{eq: equation for na}), 
we consider the solutions to the simpler equations: 
\begin{eqnarray}
D_{t}\, n^{a} = 0 \quad \textrm{and} \quad D_{i}\, n^{a} = 0, 
\end{eqnarray}  
which are explicitly given by 
\begin{eqnarray}
\partial_{\mu} n^{a} + \epsilon^{a b c} a_{\mu}^{b} n^{c} = 0\, +\, {\cal O}(u^2) .
\label{eq: NA London equation}
\end{eqnarray} 
These equations lead to the ground state solutions for the effective Hamiltonian for $n^{a}$:
\begin{eqnarray}
{\cal H}_{\mathrm{eff}} = \frac{f}{2} \left(\pi^{a}\right)^2 + \frac{1}{2} \left(D_{i}\,n^{a}\right)^2 , 
\end{eqnarray}
where the conjugate momentum is defined by $\pi^{a} = f^{-1}D_{t}\,n^{a}$.
In this article, we wish to discuss the dynamics of magnetization and spin currents 
in the boundary ferromagnetic system, which is given by the leading terms in 
the asymptotic expansions at $u \sim 0$. Hence, the higher order terms 
in the expansion with respect to $u$ are irrelevant, and we neglect them in the following. 
Dropping the ${\cal O}(u^2)$ term, 
we can easily obtain the solution to (\ref{eq: NA London equation}) for $a_{\mu}^{a}$ in terms of $n^{a}$,
\begin{eqnarray}
a_{\mu}^{a} = C_{\mu}\, n^{a} - \varepsilon^{a b c}\, n^{b} \partial_{\mu} n^{c} , 
\label{eq: relation between a and n}
\end{eqnarray}
where we have introduced a vector field $C_{\mu}$ which is arbitrary at this stage. 
This solution demonstrates the clear separation of the gauge fields: 
\begin{eqnarray}
a_{\mu}^{\parallel} = C_{\mu}\, n^{a} \quad \textrm{and} \quad a_{\mu}^{\perp} 
= -\, \varepsilon^{a b c}\, n^{b} \partial_{\mu} n^{c} .
\end{eqnarray}
The relation for the broken components, $a_{\mu}^{\perp}$, is nothing 
but a non-abelian analogue of the relation between the gauge field and 
the quantum phase of Cooper pair, $A_{\mu} = \partial_{\mu} \theta$, in superconductivity, 
and also corresponds to the Maurer-Cartan one-form of $G/H \sim SU(2)/U(1)$ in terms of 
the Nambu-Goldstone modes $n^{a}$ \cite{Leutwyler:1993gf, Roman:1999mag}.
Requiring the matching condition to the static solution (\ref{eq: static solution for time component}), 
$a_{0}^{a} = \mu_{s}\,\delta^{a 3}$ and $a_{i}^{a} = 0$ for $n^{a}=(0,0,1)$, 
the vector field $C_{\mu}$ should satisfy the condition:
\begin{eqnarray}
C_{0} = \mu_{s} \quad \textrm{and} \quad C_{i} = 0 ,
\label{eq: boundary condition for C}
\end{eqnarray}
in the static and homogeneous limit. 
Note that the relation (\ref{eq: relation between a and n}) and the ansatz 
(\ref{eq: magnetization ansatz}) do not induce  
new contributions of the scalar fields to the energy-momentum tensor $T_{MN}$ in the Einstein equations, 
and consequently the analysis in the probe approximation remain intact.    

Next, we consider the effective Yang-Mills equations, (\ref{eq: generalized time YM equation}) 
and (\ref{eq: generalized space YM equation}).  
It is not difficult to show that the relation (\ref{eq: relation between a and n}) leads to  
vanishing currents $J_{\mu}^{a}$ up to ${\cal O}(u^2)$, using the explicit form (\ref{eq: generalized current}). 
Furthermore, the ansatz for gauge fields (\ref{eq: gauge ansatz}) with $A_{u}^{a} = 0$ implies 
\begin{eqnarray}
\partial_{u} \left(u^{-1} F_{u 0}^{a}\right) = 0 , \quad \textrm{and} \quad 
\partial_{u} \left(u^{-1} f\, F_{u i}^{a}\right) = 0 + {\cal O}(u^{4}) .
\end{eqnarray}
Dropping the higher order terms 
such as ${\cal O}(u^{4})$, 
the remaining Yang-Mills equations reduce to
\begin{eqnarray}
D_{i} F_{i 0}^{a} = 0 , \quad \textrm{and} \quad D_{0} F_{0 i}^{a} + D_{j} F_{j i}^{a} = 0 .
\label{eq: boundary YM equation}
\end{eqnarray} 
From the viewpoint of the boundary theory (on the ferromagnet side), the first equation corresponds to 
a non-abelian version of Gauss's law, and the second corresponds to a non-abelian version of 
Ampere's law without source and currents, for the spin gauge fields \cite{Tatara:2017vnh}. 
Using the relation (\ref{eq: relation between a and n}), 
we obtain the $SU(2)$ field strength,\footnote{We used the relation 
$\varepsilon^{a b c} \partial_{\mu} n^{b} \partial_{\nu} n^{c} 
= n^{a} \varepsilon^{b c d} n^{b} \partial_{\mu} n^{c} \partial_{\nu} n^{d}$ due to $\sum_{a}n^{a}n^{a} = 1$.}  
\begin{eqnarray}
F_{\mu \nu}^{a} = n^{a} \left[ (\partial_{\mu} C_{\nu} - \partial_{\nu} C_{\mu}) - \varepsilon^{b c d} n^{b} \partial_{\mu} n^{c} \partial_{\nu} n^{d} \right] \equiv n^{a} f_{\mu\nu} .
\label{eq: spin EM field}
\end{eqnarray} 
Note that a component of the field strength, $f_{\mu\nu}$, parallel to the 
magnetization $n^{a}$ only remains. 
With the field strength (\ref{eq: spin EM field}), the effective Yang-Mills equations 
(\ref{eq: boundary YM equation}) and the Bianchi identity for the $SU(2)$ gauge field are  
reduced to the following equations:
\begin{eqnarray}
\partial_{\mu} f_{\mu\nu} = 0 \quad \textrm{and} \quad 
\epsilon^{\mu\nu\rho\sigma} \partial_{\nu} f_{\rho\sigma} = 0 .
\label{eq: spin Maxwell equation}
\end{eqnarray}
The above equations are the same form as the Maxwell equations, and 
the terms depending on $n^{a}$ in the gauge field $f_{\mu\nu}$ actually corresponds 
to the so-called spin electromagnetic field 
discussed in the study on ferromagnetic metals \cite{Volovik:1987wz, Tatara:2017vnh}. 
The gauge field (\ref{eq: spin EM field}) also corresponds to the unbroken $U(1)$ gauge field 
upon the symmetry breaking from $SU(2)$ to $U(1)$, with a space-dependent 
order parameter, which is frequently discussed in the context of solitonic monopoles 
in non-abelian gauge theories \cite{Harvey:1996ur}. 
Since the unbroken gauge fields in the holographic dual theory are identified 
as the (exactly) conserved currents in the boundary quantum system, 
the gauge field $C_{\mu}$ is naturally identified as the spin current with the 
polarization parallel to the magnetization $n^{a}$, which originates 
from conduction electrons.     

So far, we have obtained the relation between the gauge field $a_{\mu}^{a}$ and 
the (normalized) magnetization $n^{a}$, which implies that the gauge field dynamics can be solely 
reduced to the dynamics of the magnetization and the spin electromagnetic field $C_{\mu}$. 
Finally, we consider the remaining Yang-Mills equation in the radial $u$-direction,
in the holographic dual theory\,:  
\begin{eqnarray}
\frac{1}{\sqrt{-g}} \partial_{\mu}\left( \sqrt{-g}\, F^{\mu u\, a} \right) 
+ \epsilon^{a b c} A^{b}_{\mu} F^{\mu u\, c} = J^{u\, a} .
\label{eq: conservation law}
\end{eqnarray}
This equation is derived by the variation of the radial $u$-component of the $SU(2)$ gauge fields 
and specifies the dynamics of the gauge fields in the five-dimensional bulk; 
this equation cannot be seen in the ferromagnetic system on the boundary.  
With the ansatz (\ref{eq: magnetization ansatz}), the radial component of the current 
also vanishes ($J_{u}^{a} \equiv 0$), 
and the gauge fixing condition $A_{u}^{a} \equiv 0$ leads to the simple $SU(2)$ field strength 
$F_{\mu u}^{a} = -\,\partial_{u} A_{\mu}^{a}$ such as 
\begin{eqnarray}
&& F_{0 u}^{\parallel} = 2\, u\, a_{0}^{\parallel}(t, x) , \quad 
F_{0 u}^{\perp} = 2\, u\, a_{0}^{\perp}(t, x) , \nonumber \\ 
&& F_{i u}^{\parallel} = 2\, u\, \sigma_{s}^{\parallel}\, a_{i}^{\parallel}(t, x) , \quad  
F_{i u}^{\perp} = - 2\, u\, \sigma_{s}^{\perp}\, a_{i}^{\perp}(t, x) ,
\label{eq: u component of F}
\end{eqnarray}
where we used the ansatz (\ref{eq: gauge ansatz}) and discarded the irrelevant ${\cal O}(u^{3})$ terms.  
\begin{comment}
Inserting these forms, the equation (\ref{eq: conservation law}) can be recast as 
the following form:
\begin{eqnarray}
\partial_{0} \left(g^{00} g^{uu} F_{0 u}^{a}\right) + \partial_{i} \left(g^{ii} g^{uu} F_{i u}^{a}\right) = 0 . 
\end{eqnarray}
\end{comment}
From these forms, the second term 
in the left-hand side of (\ref{eq: conservation law}) automatically vanishes due to 
$\varepsilon^{a b c} a_{\mu}^{b} a_{\mu}^{c} = 0$.
Inserting the forms of field strength (\ref{eq: u component of F}) and 
the relation (\ref{eq: relation between a and n}), the equation can be recast as the following form: 
\begin{eqnarray}
\partial_{0} \left( C_{0}\, n^{a} - \varepsilon^{a b c} n^{b} \partial_{0} n^{c} \right) -     
\partial_{i} \left( \sigma_{s}^{\parallel}\, C_{i}\, n^{a} + \sigma_{s}^{\perp}\, 
\varepsilon^{a b c} n^{b} \partial_{i} n^{c} \right) = 0 , 
\end{eqnarray}
where the subleading terms are neglected. 
Here, we can write down the effective equation of motion for the magnetization $n^{a}$ 
in our holographic dual model:  
\begin{eqnarray}
C_{0}\, \dot{n}^{a} - \varepsilon^{a b c} n^{b} \ddot{n}^{c} -       
\sigma_{s}^{\perp}\,\varepsilon^{a b c} n^{b} \nabla^{2} n^{c} 
- \sigma_{s}^{\parallel}\,C_{i}\, \partial_{i} n^{a} = 0 ,
\end{eqnarray}
where the dot denotes the time-derivative and 
$\nabla^{2} = \partial_{i}\partial_{i}$.\footnote{A simlar analysis on effective equations 
at the linearized level in two-dimensional magnetic systems 
has been also discussed in \cite{Iqbal:2010eh}.}  
Here, we consider the condition, $\partial_{0} C_{0} - \sigma_{s}^{\parallel}\,\partial_{i} C_{i} = 0$,  
on the unbroken gauge field due to the constraint 
$\sum_{a}n^{a}n^{a} = 1$. 
This condition implies the conservation of the spin current of conduction electrons, 
which corresponds to the unbroken gauge field $C_{\mu}$, as seen below. 
Note that, since the Maxwell equations (\ref{eq: spin Maxwell equation}) 
for $C_{\mu}$ is gauge invariant, this condition can be consistently imposed as a gauge fixing condition.

Considering the matching condition (\ref{eq: boundary condition for C}), 
we decompose $C_{0}$ into $C_{0} = \mu_{s} + \tilde{C}_{0}$.  
Finally, we obtain the holographic equation for magnetization dynamics: 
\begin{eqnarray}
\mu_{s}\, \dot{n}^{a} - \varepsilon^{a b c} n^{b} \ddot{n}^{c} -  
\sigma_{s}^{\perp}\,\varepsilon^{a b c} n^{b} \nabla^{2} n^{c} 
+ \tilde{C}_{0}\, \dot{n}^{a} - \sigma_{s}^{\parallel}\, C_{i} \partial_{i} n^{a} = 0 .
\label{eq: holographic LL equation}
\end{eqnarray}
Here, we take the spin chemical potential to be 
$\mu_{s} = -\, M_{s}/\gamma$ with the magnitude of spontaneous magnetization $M_{s}$ and 
the gyromagnetic ratio $\gamma\, (>0)$,\footnote{The negative sign is introduced due to 
the negative value of the gyromagnetic ratio for electrons.}  
and also identify the spin current and spin accumulation due to conduction electrons as  
$J_{s\,i}^{\parallel} = -\, \sigma_{s}^{\parallel} C_{i}$ and $\Delta \mu_{s} = \tilde{C}_{0}$,  
using the holographic dictionary. 
Then, the holographic equation becomes the same form as 
the Landau-Lifshitz equation (without damping terms),   
\begin{eqnarray}
\frac{M_{s}}{\gamma}\,\dot{n}^{a} + \varepsilon^{a b c} n^{b} \ddot{n}^{c} = 
- \sigma_{s}^{\perp}\,\varepsilon^{a b c} n^{b} \nabla^{2} n^{c} 
+ \Delta\mu_{s}\, \dot{n}^{a} + J_{s\,i}^{\parallel}\,\partial_{i} n^{a} .
\end{eqnarray}
The last two terms in the right-hand side can be interpreted as the well-known terms 
from spin transfer torque, which describes the transfer of spin angular momentum between 
localized magnetic moments and conduction electrons \cite{Tatara:2017vnh}. Furthermore, the holographic 
Landau-Lifshitz equation (\ref{eq: holographic LL equation}) also naturally 
incorporate the spin inertia term proportional 
to the second time-derivative of the magnetization, 
which is discussed in metallic ferromagnets \cite{Kikuchi:2015cja}. 

It should be noted that the holographic Landau-Lifshitz equation 
automatically incorporates the spin transfer torque due to conduction electrons 
without introducing the corresponding fields to electrons in the dual gravitational theory. 
This is consistent with the thermodynamical results at low temperatures, which was obtained 
in the previous paper \cite{Yokoi:2015qba}.

\section{Phenomenology of Holographic Magnetization Dynamics}  
In the isotropic ferromagnets sufficiently below the Curie temperature ($T \ll T_{c}$),  
the dynamics of magnetization vector (or density of magnetic moments), $M^{a}$, is 
described by the Landau-Lifshitz equation \cite{LLequation, Lifshitz}:
\begin{eqnarray}
\frac{\partial M^{a}}{\partial t} = - \alpha\,\epsilon^{a b c} M^{b} \nabla^{2} M^{c}\quad~ 
\textrm{with}\quad~ \sum_{a=1}^{3} M^{a}M^{a} = M(T)^{2} = \textrm{const.}
\label{eq: LL equation}
\end{eqnarray}
In the following discussion, the external magnetic field and the damping term (or relaxation term) 
are ignored for simplicity.  
From the quadratic constraint, the magnetization vector can be represented as 
$M^{a}(x, t)=M(T)\,n^{a}(x,t)$ 
with the unit vector $n^{a}(x,t)$. In terms of $n^{a}(x,t)$, the Landau-Lifshitz equation becomes 
\begin{eqnarray}
M(T) \frac{\partial  n^{a}}{\partial t} = - \alpha\,M(T)^{2}\,\epsilon^{a b c} n^{b} \nabla^{2} n^{c} .
\label{eq: LL equation for n}
\end{eqnarray}
Note that the equation has two parameters, the magnitude of spontaneous magnetization, $M(T)$, at the temperature $T$, and the spin stiffness constant, $\alpha$.  

Comparing the holographic equation (\ref{eq: holographic LL equation}) with 
the Landau-Lifshitz equation (\ref{eq: LL equation for n}), 
we find that the spin chemical potential, $\mu_{s}$, in the gauge field solution 
(\ref{eq: static solution for time component}) should be proportional to the magnitude of magnetization, 
and the spin stiffness constant is given by the coeffcients $\sigma_{s}^{\perp}$ in 
the gauge field solution (\ref{eq: radial solution for A}) in the following way\,:
\begin{eqnarray}
\mu_{s} \propto - M(T) \qquad \textrm{and} 
\qquad \sigma_{s}^{\perp} \propto~ \alpha\,M(T)^{2} .
\label{eq: spin stiffness formula}
\end{eqnarray} 
In our holographic dual model, the magnitude of magnetization, $M(T)$, at the temperature $T$ 
is given by the static solution 
of the scalar field $\Phi(u)$ through the formula (\ref{eq: static solution for phi}). 
The first relation between the magnitude of magnetization and the spin chemical potential  
in ferromagnets is well-known, and frequently used as the starting point to analyze  
the various spintronic phenomena \cite{SpinCurrent}. 

Although the spin chemical potential in the equilibrium, $\mu_{s}$, is an integration constant, 
the coefficient, $\sigma_{s}^{\perp}$, is the derived quantity from the gauge field equation, 
and thus the second relation in (\ref{eq: spin stiffness formula}) on the spin stiffness constant 
is a nontrivial consequence in the holographic dual model. 
In order to obtain the coefficient, $\sigma_{s}^{\perp}$, 
we consider the linearized equation of motion for gauge fields on the background solution, 
with the static and homogeneous ansatz, $A_{i}^{\perp} = k\, G^{\perp}(u)$, 
where $k = \textrm{const}$.\footnote{The nontrivial profile $A_{x}^{\perp}(u)$ on the background 
does not contribute to the energy-momentum tensor in the Einstein equation at the linearized level.} 
Inserting this ansatz into the Yang-Mills equation 
(\ref{eq: generalized space YM equation}), we have the following linearized equation 
for $G^{\perp}(u)$:
\begin{eqnarray}
u^{3}\frac{d}{d u} \left(\frac{f(u)}{u} \left(\frac{d G^{\perp}}{d u}\right) \right) +  
\frac{\left( u\, a^{3}(u) \right)^2}{f(u)}\,G^{\perp} = 0 ,
\label{eq: holographic stiffness equation}
\end{eqnarray} 
where the metric (\ref{eq: BH metric in u}) and the $SU(2)$ gauge field 
(\ref{eq: static solution for time component}) are assumed to be the background. 
Note that this is a linear equation for $G^{\perp}$, 
and the constant $k$ is irrelevant.  
Here, we impose the first relation in (\ref{eq: spin stiffness formula}), $\mu_{s} = - M(T)/M(0)$, 
which is the magnetization normalized by the saturated magnetization, 
$M(T\,\textrm{=}\,0)$.\footnote{The proportionality 
constant is chosen for convenience in numerical calculations.}      
Using the numerical results of the holographic spontaneous 
magnetization, $M(T)$ in \cite{Yokoi:2015qba}, which is obtained using the scalar potential 
$V(|\phi|) = \lambda \left(|\phi|^2-m^2/\lambda\right)^2\!/4$ with $\lambda=1$ and $m^2=35/9$, 
we can numerically solve the equation 
(\ref{eq: holographic stiffness equation}) and obtain the asymptotic expansion (\ref{eq: radial solution for A}) near the boundary ($u \sim 0$).  
The numerical results of temperature dependences of the spin-wave stiffness, 
$D(T) \simeq \sigma_{s}^{\perp}/M(T)$, which appears in the dispersion relation of 
spin-waves, $\omega = D(T)\,k^2$, and 
the spin stiffness constant, $\alpha(T) \simeq \sigma_{s}^{\perp}/M(T)^2$, are shown in Figure 1. 

\begin{figure}[h]
\centering
\includegraphics[width=15.2cm, clip]{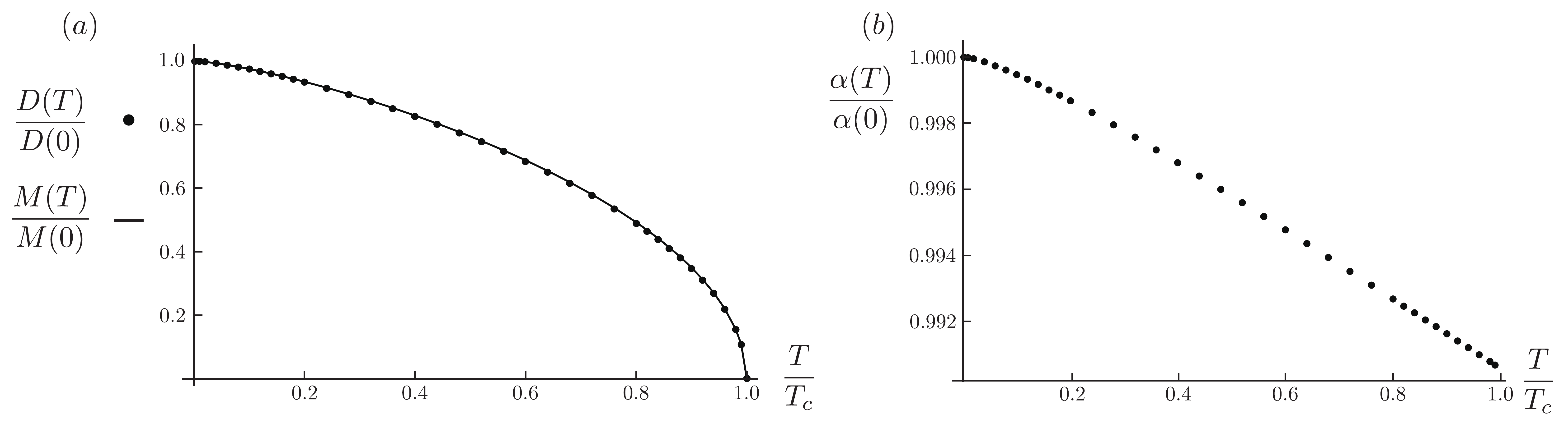}
\caption{Temperature dependence of the spin-wave stiffness is shown in Figure $(a)$:
The dots are numerical results for $D(T)/D(0)$, and the bold line is the magnetization curve, $M(T)/M(0)$.   
Temperature dependence of the spin stiffness constant, $\alpha(T)/\alpha(0)$, is shown in Figure $(b)$. 
All the results are calculated with the parameters, $\lambda=1$ and $m^2=35/9$.}
\end{figure}

The results on the spin-wave stiffness in Figure 1$(a)$ clearly show that $D(T) \propto M(T)$, 
which is consistent with the relation (\ref{eq: spin stiffness formula}) 
based on the Landau-Lifshitz equation (\ref{eq: LL equation}). 
Furthermore, the results in Figure 1$(b)$ imply the slight temperature dependence of 
the spin stiffness constant, $\alpha = \alpha(T)$, 
which can be attributed to the nonlinear spin-wave effects \cite{Atxitia}. 

A similar argument also holds for the unbroken (or parallel) component of 
the gauge fields, $A_{i}^{\parallel}$, and we can obtain the coefficient $\sigma_{s}^{\parallel}$, 
which leads to the spin torque term in the holographic Landau-Lifshitz equation 
(\ref{eq: holographic LL equation}). 
The nontrivial profile of gauge field, $A_{x}^{\parallel}(u)$, which is the parallel component to 
the spin chemical potential, $A_{0}^{\parallel}$, leads to the non-vanishing off-diagonal contribution 
in the right-hand side of the Einstein equation (\ref{eq: Einstein equation}), and thus 
induces the fluctuation of the metric $g_{t x}(u)=h_{t x}(u)/u^2$, where $h_{t x}(u)$ parameterizes 
the fluctuation finite on the boundary.  At the linearized level, 
two fluctuations, $A_{x}^{\parallel}(u)$ and $h_{t x}(u)$, form the closed equations, 
which come from the Yang-Mills equation and Einstein equation, respectively 
\cite{Hartnoll:2009sz, Herzog:2014tpa}:      
\begin{eqnarray}
&& u \frac{d}{d u} \left(\frac{f(u)}{u} \left(\frac{d A_{x}^{\parallel}}{d u}\right)\right) 
+ \left(\frac{d a^{3}(u)}{d u}\right) \frac{d}{d u} \left(u^2 h_{tx}\right) = 0 , \\
&& u^{-2} \frac{d}{d u} \left(u^2 h_{tx}\right) + 2 \left(\frac{d a^{3}(u)}{d u}\right) A_{x}^{\parallel} = 0 .
\end{eqnarray}
Deleting the metric fluctuation, we can obtain the equation for $G^{\parallel}(u)$:
\begin{eqnarray}
u \frac{d}{d u} \left(\frac{f(u)}{u} \left(\frac{d G^{\parallel}}{d u}\right)\right) 
-  2 u^{2} \left(\frac{d a^{3}(u)}{d u}\right)^2 G^{\parallel} = 0 .
\end{eqnarray}    
We can numerically solve the equation, and obtain the coefficient $\sigma_{s}^{\parallel}$ from  
the asymptotic expansion of the solution in (\ref{eq: radial solution for A}).     
The resulting temperature dependence of the spin torque coefficient, 
$\tau_{s}(T) = \sigma_{s}^{\parallel}/M(T)$, is shown in Figure 2. 

\begin{figure}[h]
\centering 
\includegraphics[width=9cm, clip]{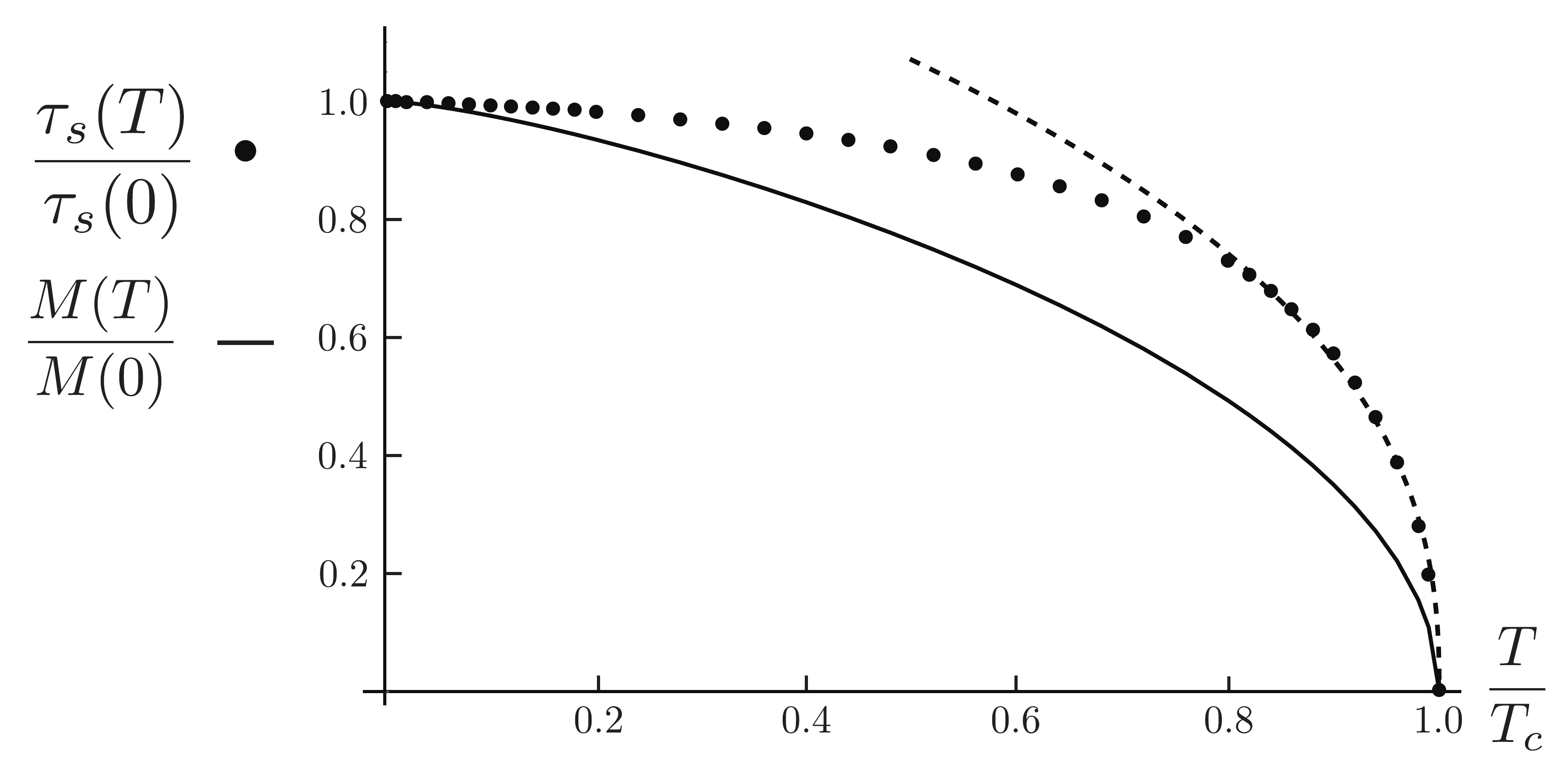} 
\caption{Temperature dependence of the spin torque coefficient: 
The dots are numerical results for $\tau_{s}(T)/\tau_{s}(0)$, and the bold line is 
the magnetization curve $M(T)/M(0)$. The dashed line is the fitting curve,  
$\tau_{s}(T)/\tau_{s}(0) = c \left(1-T/T_{c}\right)^{2/5}$ with $c\simeq1.41$, near the Curie temperature.} 
\end{figure}

The results on the magnitude of the spin transfer torque, $\tau_{s}(T)$,  
show that the spin torque effect is approximately constant at low temperatures 
(in comparison with magnetization curve), 
and is vanishing towards the Curie temperature as $\tau_{s}(T) \propto (1-T/T_{c})^{2/5}$. 
This property at low temperatures is consistent with the phenomenological form of 
the spin transfer torque, $(\bm{J}_{s}^{\parallel}\cdot\bm{\nabla} n^{a})/M(T)$, 
whose magnitude is independent of the norm of magnetization due to $|\bm{J}_{s}^{\parallel}| \propto M(T)$ at the leading order \cite{Tatara:2017vnh}. 
In addition, the finite spin torque coefficient is a consequence of the \textit{both} 
fluctuations of the gauge field and metric. 
In accordance with the holographic dictionary \cite{Hartnoll:2009sz}, the metric fluctuation $h_{t x}$ 
corresponds to the temperature gradient, $\nabla_{x} T/T$, in the ferromagnetic system.  
This calculation implies that 
the effect of spin transfer torque appears only in the nonequilibrium situations, 
where spin transfer is accompanied by heat (or entropy) transfer.

\section{Summary and Discussion}
We have discussed a novel approach to understand magnetization dynamics in ferromagnets 
using the holographic realization of ferromagnetic systems. 
The Landau-Lifshitz equation describing magnetization dynamics was 
derived from the Yang-Mills-Higgs equations in the dual gravitational theory. 
This holographic Landau-Lifshitz equation automatically incorporates 
not only the exchange interaction 
but also the spin transfer torque effect due to conduction electrons.
Furthermore, we numerically investigated the temperature dependences of 
the spin-wave stiffness and the magnitude of spin transfer torque in the 
holographic dual theory, and  
the results obtained so far are consistent with the known properties of 
magnetization dynamics in ferromagnets with conduction electrons.  

This holographic approach to magnetization dynamics can be applied to 
more generic situations. 
For instance, the holographic Landau-Lifshitz equation can incorporate the damping term 
by considering more generic metric fluctuations, which correspond to phonon dynamics 
in the boundary ferromagnets.   
Moreover, the holographic dual theory may provide geometric approaches  
to spin caloritronics \cite{SpinCaloritronics}, 
where magnetization dynamics is considered under temperature gradients, from 
higher dimensional perspectives.  
We thus believe that the holographic approach provides useful tools 
to analyze nonequilibrium and nonlinear dynamics of magnetization in ferromagnets,  
and also leads to new perspectives in spintronics from gravitational physics.

\vspace*{0.5cm}
\noindent
{\large {\bf Acknowledgement}}

\noindent
The authors thank M. Ishihara for collaboration at the early stage of this work, 
and also K. Harii and Y. Oikawa for useful discussions. 
The works of N. Y. and E. S. were supported in part by Grant-in Aid for Scientific Research on 
Innovative Areas "Nano Spin Conversion Science" (26103005), and  
the work of K. S. was supported in part by JSPS KAKENHI Grant No. 
JP17H06460. 
The works of N. Y. and E. S. were supported in part by ERATO, JST.

\end{document}